\documentclass[sigconf] {acmart}
\renewcommand\footnotetextcopyrightpermission[1]{} 
\usepackage{graphicx} 

\begin{document}

\title{Human Latency Conversational Turns for Spoken Avatar Systems}

\author{Derek Jacoby}
\affiliation{%
  \institution{University of Victoria}
  \city{Victoria, BC}
  \country{Canada}
}
\author{Tianyi Zhang}
\affiliation{%
  \institution{University of Victoria}
  \city{Victoria, BC}
  \country{Canada}
}
\author{Aanchan Mohan}
\affiliation{%
  \institution{Northeastern University}
  \city{Vancouver, BC}
  \country{Canada}
}
\author{Yvonne Coady}
\affiliation{%
  \institution{Northeastern University}
  \city{Vancouver, BC}
  \country{Canada}
}

\maketitle

\section*{Abstract}
A problem with many current Large Language Model (LLM) driven spoken dialogues is the response time. Some efforts such as Groq address this issue by lightning fast processing of the LLM, but we know from the cognitive psychology literature that in human-to-human dialogue often responses occur prior to the speaker completing their utterance. No amount of delay for LLM processing is acceptable if we wish to maintain human dialogue latencies. 

In this paper, we discuss methods for understanding an utterance in close to real time and generating a response so that the system can comply with human-level conversational turn delays. This means that the information content of the final part of the speaker's utterance is lost to the LLM.  Using the Google NaturalQuestions (NQ) database, our results show GPT-4 can effectively fill in missing context from a dropped word at the end of a question over
60\% of the time.  
These results indicate that a simple classifier could be used to determine whether a question is semantically complete, or requires a filler phrase to allow a response to be generated within human dialogue time constraints.

\section{Introduction}

In this paper, we would like to explore some methods of evaluating spoken language avatar dialogue systems with respect to human-to-human dialogues. We will discuss response quality, and some mechanisms of evaluating that quality, but our ultimate focus will be on latencies in human dialogues and corresponding latencies in automated systems. In a 2009 study on latencies in conversational turn taking \cite{stivers_universals_2009} it was found that on average in English there is a 239 msec delay between the original interlocutor in a dialogue and the start of the answering utterance. The variance is quite high, though, with one standard deviation in response time being 519 msec. So human dialogue expectations are that responses occur between -280 and +758 msecs from the end of an utterance, on average. In other languages, these expectations of response time are even more challenging for automated systems to meet, for instance in Japanese the average response is only 7 msec after the original speaker finishes speaking.

The current standard for architectures for automated dialogue systems is one where the speech recognition pass begins when the utterance ends, and then the recognized speech is passed to a Large Language Model (LLM) which composes the response to be sent to a synthetic speech generation module. This architecture is completely unable to conform to the latency expectations in human dialogues. This serial processing of responses is also counter to the means by which humans process speech. Humans form an understanding in real time and engage in a number of complex turn-taking behaviours to negotiate when to take control of the conversation~\cite{lunsford_measuring_2016}. In this paper we will not delve too deeply into those cues, such as eye contact and non-verbal utterances, which might signal a desire to begin to speak, but we will concentrate on the timings when it is turn for the system to speak. Violating the conversational expectations of human to human dialogue detracts from the perception of naturalness and engagement in human-machine spoken dialogues, although users do have some tolerance for longer delays than would be acceptable in a human conversational partner\cite{peng_understanding_2020}.

There is a deep literature on real-time speech recognition systems which will be briefly reviewed in the related work section. In general, these systems lose some accuracy in comparison with speech recognition systems that can search from both ends of an utterance in generating its recognition candidates. We will discuss semantic redundancy in dialogue utterances, though, and provide experimental evidence that in many cases the loss of recognition accuracy is not a detriment to the quality of spoken avatar responses. We will also discuss heuristics for the use of filler phrases when the construction of an utterance leads the avatar system to suspect that critical information is being withheld until the end of an utterance. Finally, we will speculate on the impact of these conversational behaviours in the construction of spoken dialogues in the construction of an avatar that we are in the midst of building to support engaging the general public in a museum experience.

To summarize, the realization that this work is being driven from is that in American English a conversational turn generally begins from 239 msec before the first speaker is done speaking to about 758 msec after. There have been attempts to on each dialogue turn use a filler phrase of some sort to start the automated dialog response while the system runs the speech recognition and generates the response, but this gets unnatural quickly. Our current attempt is to take a question answering dataset, chop the final 1, 2 or 3 words off the initial utterance and generate a response, use an automated framework to judge the quality of the response, and then model how much accuracy we lost on the responses by lopping the initial question off early. Next, to use the instances that lost accuracy to form a binary classifier so that when we are part way through the question we can take the classifier output to determine whether we should respond with a filler phrase and then process the entire question (late informative questions) or skip the filler phrase and process an answer based on the truncated question (late uninformative questions). The hypothesis being that by classifying and making a decision on whether to complete processing that we can recover answer quality with a minimum of use of filler phrases by taking advantage of a statistical understanding of sentence structure to determine the likelihood of a twist at the end of the sentence.

We are taking a statistical pattern recognition approach to this problem, there have also been linguistic processing approaches related to our efforts, but they require a deeper semantic modeling of the input phrases which seem to require more hand-processed rules to execute~\cite{zhou_online_2022}. This rule-based versus statistical approach in some ways mirrors ongoing theoretical positions within the linguistics community which we will elaborate upon in related work.

\section{Background and Related Work}
In this review we are going to examine different methods for assessing machine dialogues including some discussion of data sets. Next we will review some latencies and factors affecting latency in human-to-human dialogues, this will also include a discussion of underlying neural correlates of speech in humans. Finally we will identify some of the limiting factors in the latency of machine dialogues, and detail some previous attempts to overcome these barriers.

\subsection{Assessing machine dialogues}
We later propose some methods for speeding up machine dialogues, but it is not possible to do this without first carefully considering the impact on the quality of those dialogues. There are a variety of ways of assessing dialogue quality, and this has an interplay with the type of dataset used for that assessment. In our case, we will be mostly using the Google NaturalQuestions database \cite{kwiatkowski_natural_2019} which has large number of factual questions (drawn from Wikipedia) with short, long, and yes/no answers supplied by human annotators. This gold standard approach is effective in that a human annotator supplies the ground truth, but this annotation is expensive. This particular data set has, in some ways, been supplanted by more challenging tasks \cite{rosset_researchy_2024} but we have elected to set a baseline on our approach before going to a more variable dataset.

Some approaches have tried to automate the comparison of human responses to the automatically generated responses\cite{alberti_bert_2019}. The other alternatives are reference-free approaches, for instance LLM-eval \cite{lin_llm-eval_2023} is a common reference-free approach that uses a multi-dimensional set of automated metrics to evaluate dialogue quality. We have chosen to use a method that relies on a human annotated gold standard, but uses an LLM to calculate a semantic similarity score, semScore, to that gold standard~\cite{zheng_judging_2023}. This method, LLM-as-judge, allows us to compare against the human annotated standard, and then from there against our questions manipulations in each case giving a 0 to 1 score for how semantically similar the two answers are based on a cosine distance measurement.

\subsection{Latencies in Human-to-Human Dialogue}
The analysis of human-to-human dialogue latencies is complex both in terms of which language is under discussion, and what the intent of the dialogues are \cite{stivers_universals_2009}. A more in-depth analysis of dialogue intents, and the means of grounding dialogues, is found in a book chapter on Cognitive Mathematics, which also does a good job of describing the types of dialogues \cite{danesi_understanding_2022}. In our case, we are focusing on information seeking dialogues.

Even the measurement of turn taking has some complexities, and are referred as overlaps (where the reply begins before the prior utterance is finished) and gaps (where there is some silence between turns)~\cite{lunsford_measuring_2016}. Some Scandinavian studies have re-enforced our ideas of the importance of smooth turn taking in establishing pleasing machine dialogues in both a paper \cite{jonsdottir_learning_2008} and a doctoral thesis~\cite{hjalmarsson_human_2010}.

\subsection{Spoken Dialogue Theories}

As far back as the mid-1990's researchers began studying the nature of human-machine spoken dialogues~\cite{johnstone_there_1995} Given the technology of the day, this involved Wizard of Oz studies, with a human with a script simulating the machine side of the dialogue. A fundamental realization of that, and most subsequent studies, is that human-machine dialogues are simply different that human-to-human dialogues. The expectations are different and the interlocutory acts are different. We maintain that this is an artifact of the state of the technology and that as this technology improves that human-machine spoken dialogue will more closely resemble the dialogue between people. Of course, some linguistic theorists would ardently disagree, and in fact would maintain that today's statistically-generated dialogue behaviours are fundamentally different to human language mechanisms, and thus a natural dialogue is largely impossible~\cite{chomsky_noam_2023}. In fact, the very existence of Large Language Models in some ways threatens the Chomsky view that language depends on innate structures, and is not simply an emergent property of a probabilistic system~\cite{piantadosi_modern_2023}.

There is a class of sentence that is studied which is known as a sluiced sentence. This means that there is a portion of the sentence that is ommitted and filled in from context. This type of sentence is very common where the context exists in a proximal sentence and requires additional processing by the person listening to fill in the missing context~\cite{clifton_comprehension_1998}. One can consider the questions where we are truncating the last words as a form of sluiced sentence. If the context is too hard to recover, however, processing of the question will fail.

\subsection{Latencies in Machine Dialogues}
Most spoken dialogue avatar systems attempt to simply process as quickly as possible in order to generate a response within an acceptable amount of time for the user. The elements of that processing generally include the speech recognition pass (which for the best recognition quality cannot start until the utterance is complete), followed by a language understanding pass (generally an LLM currently) to produce the reply, followed by a Test-to-Speech (TTS) system to generate the output audio. Each of these steps take time.

Let us first consider the recognition speed. The OpenAI whisper speech recognition engine has made a large impact on the community, but generally provides only offline recognition - meaning that the entire utterance is processed at once. Some efforts have been made to perform online, or ongoing, recognition using whisper~\cite{lyu_real-time_2024}. In many situations it would be ideal to use a cloud service for the speech recognition, but there are large differences in the latency of the offerings by different providers. Microsoft tends to be the faster in providing both initial hypotheses and final recognitions~\cite{addlesee_comprehensive_2020}. In general on Azure, the first hypotheses come back within 150ms and those hypotheses then to be 95\% stable within 500 msec. So we could depend on final recognitions within approximately 650ms from the end of the utterance. Google and IBM on their cloud services would take on the order of two seconds to produce their final recognitions.

The next item to look at is the response generation. LLMs can very widely in their latencies, but one of the fastest is the new Groq API. There are some restrictions, but on the open source models that they have optimized 240 tokens per second is not unreasonable with only a few msec of initial latency in the response whereas the speed of generation on OpenAI is closer to 94 ms per token~\cite{noauthor_artificialanalysisai_2024, noauthor_gpt-35_2023}. Given that the responses in this study range from about 20 tokens to about 60 tokens in length, the response generation time in the worst cases will be approximately 250 msec on Groq or 650 msec on OpenAI.

On the speech production side, numbers are very situation dependent but we have found that between 80 and 100 msec response times are a reasonable estimate, and this aligns with other experiences in the literature~\cite{ahmad_enhancing_2024}.

So an overall response time, independent of network delays, is likely in the 1000 to 1500 msec range for most questions.

\section{Methodology and Experimental Framework}
This section gives an overview of the experimental setup for this work. Our goal is to simulate two cases in human dialogue when it comes to question answering, namely the late informative question scenario and late uniformative question scenario. We would then like to understand the impact this has on responses generated by an LLM. This section first briefly describes the Google NaturalQuestions (NQ) dataset, followed by our setup to mimic the two stated question answering scenarios. Furthermore, we describe the process of scoring similarities between responses generated by the LLM for the different scenarios compared to the ground-truth and the baseline replies generated by the LLM.
\subsection{Google NaturalQuestions Dataset}
The Google NaturalQuestions(NQ) dataset~\cite{kwiatkowski_natural_2019} as mentioned before consists of answers to a large number of factual questions. The dataset consists of 307,383 training examples, 7830 development examples which contain human annotations and 7842 test examples. For our experimental study we chose the first 1,000 examples in the development set as our subset for the purposes of this experimental study. Each example consists of a question along with a long answer that is generated by human annotators which we treat as the gold-standard answer (which will be referred to as `ref`). We dropped those questions which were negative controls (meaning that they were designed not to be answerable only from the question text). We also intentionally did not use the context text from the dataset, instead preferring to use only the background training data from ChatGPT to provide context for the question. This was to not favorably bias the correct answering of questions from the context provided.

\subsection{Responses from Large Language Models}
In our experimental study we used ChatGPT~\cite{brown_language_2020} as our LLM to generate responses for each of the 1,000 example questions taken from the NQ dataset. For each question in our subset, the response that that is generated our LLM is referred to as `res-0' or response-0. This refers to responses without any sentence truncation in the original question. This response simulates our late informative response scenario in natural dialogue, where a speaker giving the answer listens to the entire phrase before responding.

In order to simulate the long uninformative response scenario, where a speaker giving the answer starts before the question is finished being said.  The typical speaking rate for English is 4 syllables per second~\cite{cruttenden_gimsons_2014}. It is well known that some of the most common words in the English language are between 1-7 syllables in the length with a large number of words being not more than 3-4 syllables long. Considering this, we consider two subsets of our 1,000 example subset. In the first subset, we take each of the examples and remove the last word in the each question and record the responses from our LLM which we refer to as `res-1'. This simulates removal of approximately between 500ms-1sec. of speech audio. Similarly, in our second subset, we truncate the last two words in each question and record the responses from our LLM which we refer to as `res-2'. This simulates the removal of approximately 1sec or more of speech audio. In our third subset we truncate the last 3 words and we refer to this as `res-3'.

\subsection{Scoring similarity between responses}
In order to evaluate the quality of responses we use SEMSCORE~\cite{aynetdinov_semscore_2024} a measure of semantic textual similarity. In the calculation of the SEMSCORE, the ground truth response to a question, the target response, and the response from the LLM, the so-called model response are both converted to sentence embeddings using a sentence transformer model `all-mpnet-base-v2'~\cite{reimers_sentence-bert_2019}. The SEMSCORE then consists of calculating the cosine similarity between the ground truth target response embedding and the embedding corresponding to the LLM response. The value of the SEMSCORE lies between [-1,1]. If the scores are close to 1, then the two sentences are considered semantically similar. Negative values imply semantically opposite sentences.

\section{Results}
\begin{figure}[ht]
  \centering
    \includegraphics[width=\columnwidth]{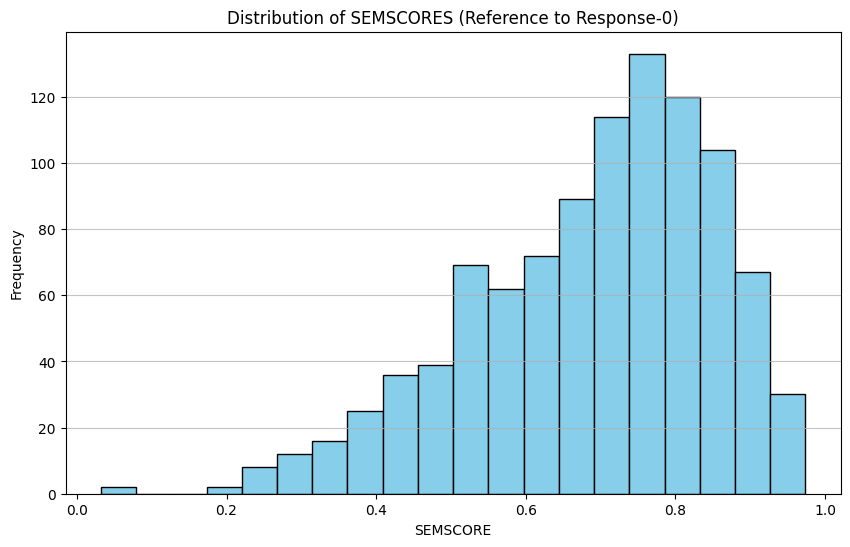}
  \caption{Histogram of SEMSCORES for LLM returned responses against the ground truth reference responses}
  \label{fig:res0-ref}
\end{figure}

This section describes our experiments. In order to understand how closely the responses from the LLM match to the ground truth responses, the SEMSCORE is first calculated between the responses received from the LLM (`res-0') and the human annotated ground truth responses (`ref`). Figure~\ref{fig:res0-ref} shows a histogram plot of the SEMSCORES from our set `res-0' scored against the ground truth responses. The LLM used to generate responses in this figure is ChatGPT with the GPT-4 model. The histogram is seen to provide a central tendency of scores closer to a mean of 0.68 to indicate that the returned responses and the ground truth responses bear semantically similarity. The standard deviation is 0.16 with the minimum SEMSCORE being 0.03, and the maximum SEMSCORE being 0.97. The 75th percentile of these scores is 0.81.

In order to understand the impact of word truncation, our next experiment looks at the distribution of scores for examples whose complete responses from the LLM very closely matched the human annotated ground-truth responses. For this we selected those `res-0' examples whose `res-0' vs `reference' scores were above the 75th percentile score of 0.81. We then plotted the SEMSCOREs for these examples comparing : 
\begin{itemize}
\item The original response from the LLM (`res-0') and its similarity compared to the response obtained when the last word was removed from the original question (`res-1').
\item The original response from the LLM (`res-0') and its similarity compared to the response obtained when the last two words were removed from the original question (`res-2').
\item The original response from the LLM (`res-0') and its similarity compared to the response obtained when the last three words were removed from the original question (`res-3').
\end{itemize}

\begin{figure}[ht]
  \centering
    \includegraphics[width=\columnwidth]{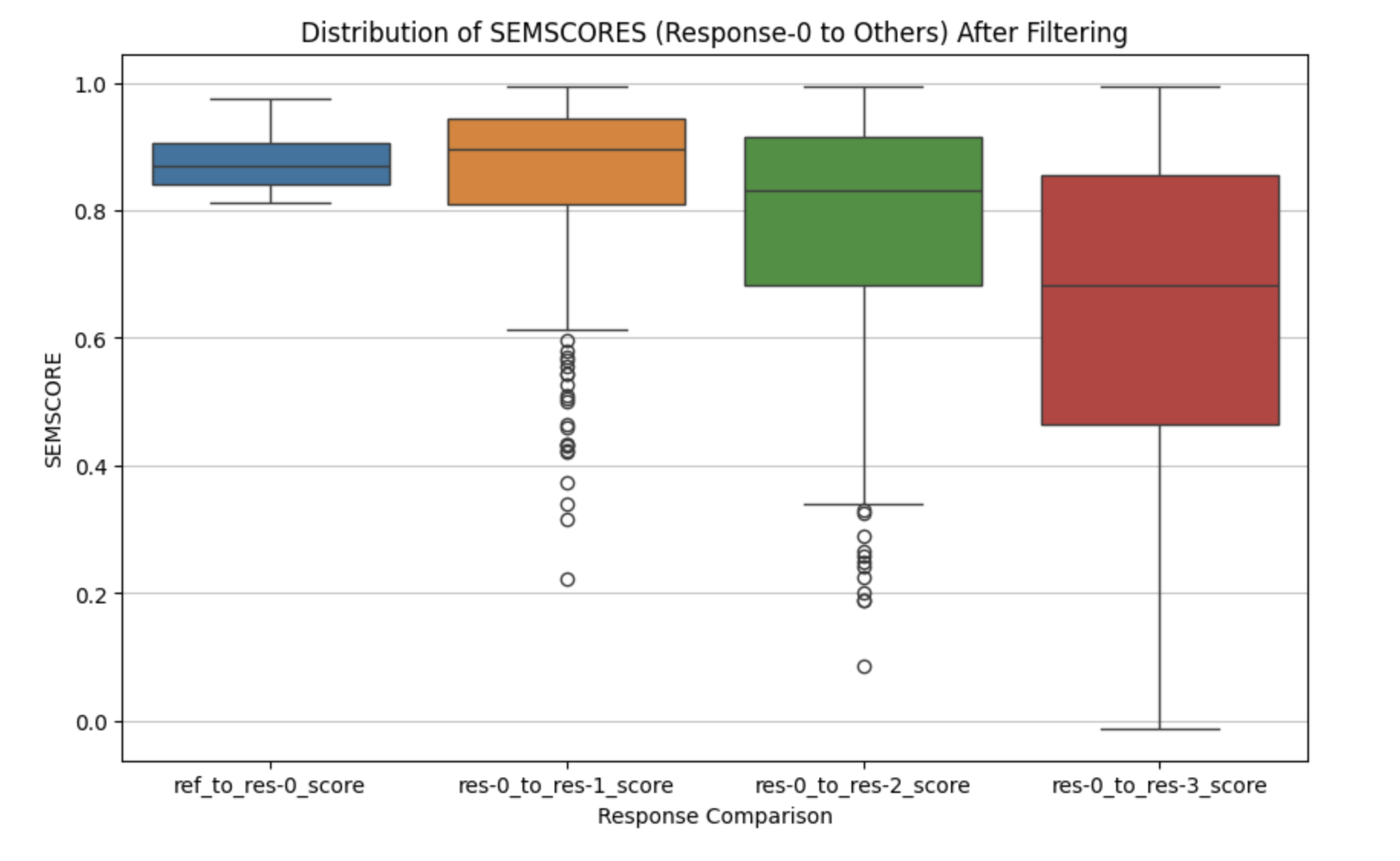}
  \caption{Box and whisker plots of SEMSCORES for the similarity of `res-0' vs truncated utterances, specifically those utterances whose `res-0' scores when compared to the ground truth were above the 75th percentile score of 0.81}.
  \label{fig:res0-to-other-res}
\end{figure}

\begin{figure}[ht]
  \centering
    \includegraphics[width=\columnwidth]{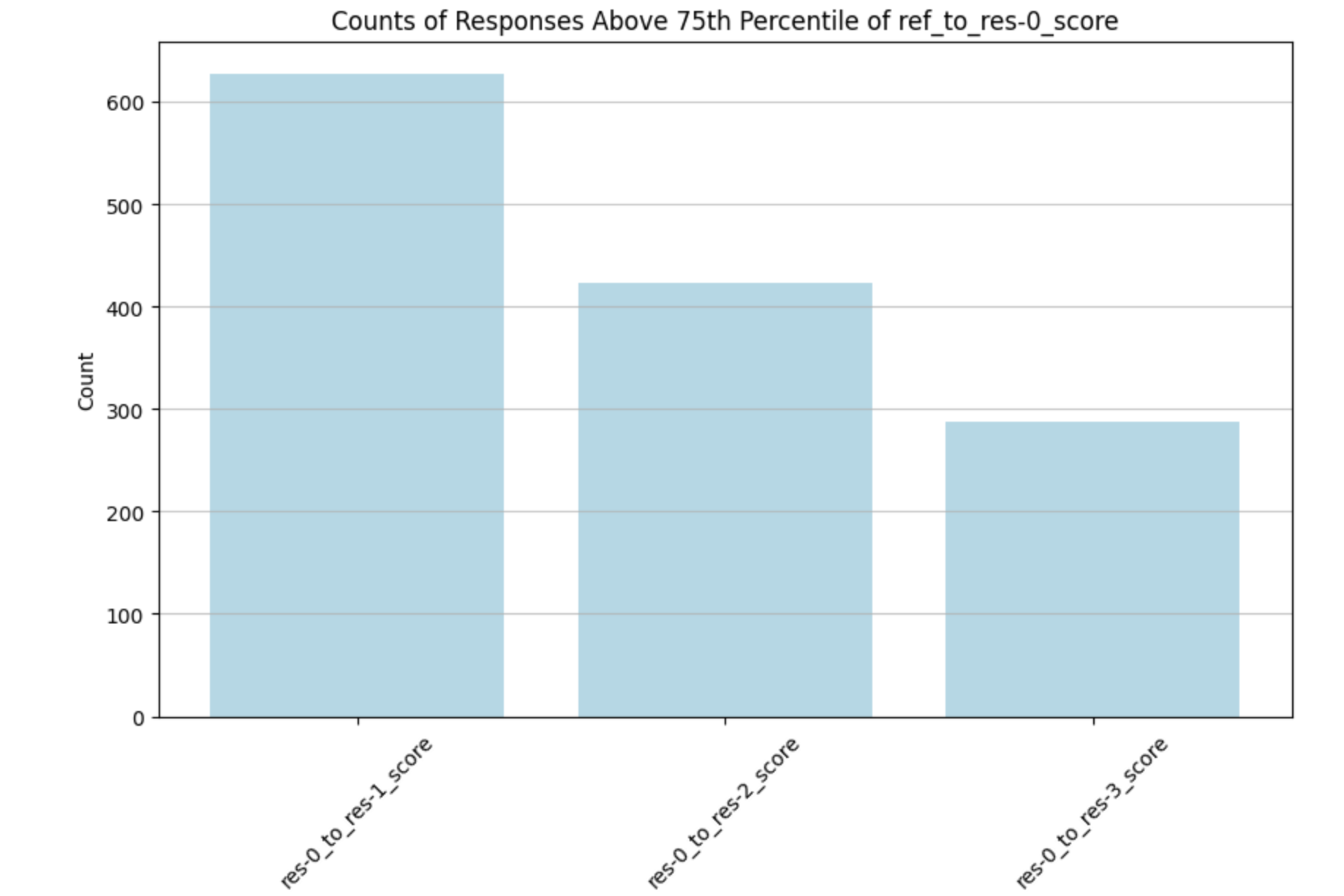}
  \caption{Number of questions (out of 1000) where the response was rated within 75th percentile of untruncated responses}.
  \label{fig:counts}
\end{figure}

The distribution of these scores for these specific examples are captured in the box and whisker plots in Figure~\ref{fig:res0-to-other-res}. 
The counts of scores rated above 75th percentile for each truncation condition are shown in Figure~\ref{fig:counts}.

\section{Discussion}

We have presented data showing that the use of state of the art LLM's such as GPT-4 can quite effectively fill in missing context from dropped words in a question. In the case of a single dropped word, this results in little impact on the quality of the response over 60\% of the time. In pursuit of human latencies in spoken avatar dialogues, we are in the midst of taking advantage of this in the preparation of filler phrases which will allow the avatar to respond within latency expectations if it is still processing. For instance, in the question "How do I get from New York to Chicago?" plainly the question cannot be answered until after the entire question is uttered. This is an example of a late informative question, and a human conversational partner might answer "Well, the way I would go would be to take interstate..." with the first part of that response ("Well, the way I would go") essentially playing the role of a filler phrase while the rest of the response is composed. If the question were asked as "I want to drive from New York to Chicago, can you give me directions?" then no filler phrase would be needed.

In some sense, the capabilities of the language model to fill in missing context (deal with sluiced sentences) determines how often filler phrases will be needed to maintain appropriate conversational norms. That GPT-4 is so superior to previous language model variants allows us to rely less on context directly in the questions and instead depend on the LLM to fill in the gaps. What this gap filling cannot do, though, is allow the avatar to engage in other conversation mediating activities. A human conversational partner will make eye contact, nod, make non-verbal utterances of agreement in an attempt to re-assure the speaker that their message is being received. It is only if the speech recognizer is running on the fly and the avatar paying attention to the semantics of the question that these types of conversation reinforcing behaviors can be produced. The production of filler phrases and adherence to human conversational norms is only one of the small benefits of processing, as humans do, while the conversational turn is ongoing.

The use of heuristics, like filler phrases and rules for semantic non-interrupting responses, can be thought of as something of a bridge between purely statistical language learning in the LLM, and higher level language rules that are so prominent in Chomsky's approach to language~\cite{chomsky_noam_2023}.

\section{Conclusions and Future Work}

This work is ongoing as we develop an avatar for responding to questions about the environment that will be on display in a museum setting. Before we release the avatar we will need to move our current text-based investigations into spoken language inputs and responses. There are several datasets of spoken language questions, but we intend to develop our own that are in-domain with the environmental questions that we expect our users to ask.

The work to build a classifier to allow us to determine on an ongoing basis as the user speaks whether the question will be late informative is work that is upcoming. Specifically, we intend to use the questions in our dataset that were labeled late informative through the use of the semScore measure differing from complete questions to train a classifier. This will allow us to develop a score that tends towards 1 for identifying when a question is semantically complete, and as we pass a cutoff that we define we can then prepare the response. In those cases where the question ends without a response being ready we will have used the intermediate recognitions (and semantic understanding) to generate an appropriate filler phrase with as much specificity as the question allows.

Eventually, we also expect to use these intermediate semantics to introduce dialogue reinforcing behaviors. This is particularly the case as our graphical avatar is produced to go along with our responses.

The mix of systems latency questions and user expectations make this an exciting time to continue to craft machine conversational systems that meet user expectations of human conversational partners.


\bibliographystyle{ACM-Reference-Format}
\bibliography{UIST2024}


\begin{thebibliography}{25}


\ifx \showCODEN    \undefined \def \showCODEN     #1{\unskip}     \fi
\ifx \showDOI      \undefined \def \showDOI       #1{#1}\fi
\ifx \showISBNx    \undefined \def \showISBNx     #1{\unskip}     \fi
\ifx \showISBNxiii \undefined \def \showISBNxiii  #1{\unskip}     \fi
\ifx \showISSN     \undefined \def \showISSN      #1{\unskip}     \fi
\ifx \showLCCN     \undefined \def \showLCCN      #1{\unskip}     \fi
\ifx \shownote     \undefined \def \shownote      #1{#1}          \fi
\ifx \showarticletitle \undefined \def \showarticletitle #1{#1}   \fi
\ifx \showURL      \undefined \def \showURL       {\relax}        \fi
\providecommand\bibfield[2]{#2}
\providecommand\bibinfo[2]{#2}
\providecommand\natexlab[1]{#1}
\providecommand\showeprint[2][]{arXiv:#2}

\bibitem[noa(2023)]%
        {noauthor_gpt-35_2023}
 \bibinfo{year}{2023}\natexlab{}.
\newblock \bibinfo{title}{{GPT}-3.5 and {GPT}-4 response times}.
\newblock
\newblock
\urldef\tempurl%
\url{https://www.taivo.ai/__gpt-3-5-and-gpt-4-response-times/}
\showURL{%
\tempurl}


\bibitem[noa(2024)]%
        {noauthor_artificialanalysisai_2024}
 \bibinfo{year}{2024}\natexlab{}.
\newblock \bibinfo{title}{{ArtificialAnalysis}.ai {LLM} {Benchmark} {Doubles} {Axis} - {Groq}}.
\newblock
\newblock
\urldef\tempurl%
\url{https://wow.groq.com/artificialanalysis-ai-llm-benchmark-doubles-axis-to-fit-new-groq-lpu-inference-engine-performance-results/}
\showURL{%
\tempurl}
\newblock
\shownote{Section: Blog}.


\bibitem[Addlesee et~al\mbox{.}(2020)]%
        {addlesee_comprehensive_2020}
\bibfield{author}{\bibinfo{person}{Angus Addlesee}, \bibinfo{person}{Yanchao Yu}, {and} \bibinfo{person}{Arash Eshghi}.} \bibinfo{year}{2020}\natexlab{}.
\newblock \showarticletitle{A {Comprehensive} {Evaluation} of {Incremental} {Speech} {Recognition} and {Diarization} for {Conversational} {AI}}. In \bibinfo{booktitle}{\emph{Proceedings of the 28th {International} {Conference} on {Computational} {Linguistics}}}, \bibfield{editor}{\bibinfo{person}{Donia Scott}, \bibinfo{person}{Nuria Bel}, {and} \bibinfo{person}{Chengqing Zong}} (Eds.). \bibinfo{publisher}{International Committee on Computational Linguistics}, \bibinfo{address}{Barcelona, Spain (Online)}, \bibinfo{pages}{3492--3503}.
\newblock
\urldef\tempurl%
\url{https://doi.org/10.18653/v1/2020.coling-main.312}
\showDOI{\tempurl}


\bibitem[Ahmad(2024)]%
        {ahmad_enhancing_2024}
\bibfield{author}{\bibinfo{person}{Syed~Rameel Ahmad}.} \bibinfo{year}{2024}\natexlab{}.
\newblock \bibinfo{title}{Enhancing {Multilingual} {Information} {Retrieval} in {Mixed} {Human} {Resources} {Environments}: {A} {RAG} {Model} {Implementation} for {Multicultural} {Enterprise}}.
\newblock
\newblock
\urldef\tempurl%
\url{https://doi.org/10.48550/arXiv.2401.01511}
\showDOI{\tempurl}
\newblock
\shownote{arXiv:2401.01511 [cs]}.


\bibitem[Alberti et~al\mbox{.}(2019)]%
        {alberti_bert_2019}
\bibfield{author}{\bibinfo{person}{Chris Alberti}, \bibinfo{person}{Kenton Lee}, {and} \bibinfo{person}{Michael Collins}.} \bibinfo{year}{2019}\natexlab{}.
\newblock \bibinfo{title}{A {BERT} {Baseline} for the {Natural} {Questions}}.
\newblock
\newblock
\urldef\tempurl%
\url{http://arxiv.org/abs/1901.08634}
\showURL{%
\tempurl}
\newblock
\shownote{arXiv:1901.08634 [cs]}.


\bibitem[Aynetdinov and Akbik(2024)]%
        {aynetdinov_semscore_2024}
\bibfield{author}{\bibinfo{person}{Ansar Aynetdinov} {and} \bibinfo{person}{Alan Akbik}.} \bibinfo{year}{2024}\natexlab{}.
\newblock \bibinfo{title}{{SemScore}: {Automated} {Evaluation} of {Instruction}-{Tuned} {LLMs} based on {Semantic} {Textual} {Similarity}}.
\newblock
\newblock
\urldef\tempurl%
\url{http://arxiv.org/abs/2401.17072}
\showURL{%
\tempurl}
\newblock
\shownote{arXiv:2401.17072 [cs]}.


\bibitem[Brown et~al\mbox{.}(2020)]%
        {brown_language_2020}
\bibfield{author}{\bibinfo{person}{Tom~B. Brown}, \bibinfo{person}{Benjamin Mann}, \bibinfo{person}{Nick Ryder}, \bibinfo{person}{Melanie Subbiah}, \bibinfo{person}{Jared Kaplan}, \bibinfo{person}{Prafulla Dhariwal}, \bibinfo{person}{Arvind Neelakantan}, \bibinfo{person}{Pranav Shyam}, \bibinfo{person}{Girish Sastry}, \bibinfo{person}{Amanda Askell}, \bibinfo{person}{Sandhini Agarwal}, \bibinfo{person}{Ariel Herbert-Voss}, \bibinfo{person}{Gretchen Krueger}, \bibinfo{person}{Tom Henighan}, \bibinfo{person}{Rewon Child}, \bibinfo{person}{Aditya Ramesh}, \bibinfo{person}{Daniel~M. Ziegler}, \bibinfo{person}{Jeffrey Wu}, \bibinfo{person}{Clemens Winter}, \bibinfo{person}{Christopher Hesse}, \bibinfo{person}{Mark Chen}, \bibinfo{person}{Eric Sigler}, \bibinfo{person}{Mateusz Litwin}, \bibinfo{person}{Scott Gray}, \bibinfo{person}{Benjamin Chess}, \bibinfo{person}{Jack Clark}, \bibinfo{person}{Christopher Berner}, \bibinfo{person}{Sam McCandlish}, \bibinfo{person}{Alec Radford}, \bibinfo{person}{Ilya Sutskever},
  {and} \bibinfo{person}{Dario Amodei}.} \bibinfo{year}{2020}\natexlab{}.
\newblock \bibinfo{title}{Language {Models} are {Few}-{Shot} {Learners}}.
\newblock
\newblock
\urldef\tempurl%
\url{https://doi.org/10.48550/arXiv.2005.14165}
\showDOI{\tempurl}
\newblock
\shownote{arXiv:2005.14165 [cs]}.


\bibitem[Chomsky et~al\mbox{.}(2023)]%
        {chomsky_noam_2023}
\bibfield{author}{\bibinfo{person}{Noam Chomsky}, \bibinfo{person}{Ian Roberts}, {and} \bibinfo{person}{Jeffrey Watumull}.} \bibinfo{year}{2023}\natexlab{}.
\newblock \showarticletitle{Noam chomsky: {The} false promise of chatgpt}.
\newblock \bibinfo{journal}{\emph{The New York Times}}  \bibinfo{volume}{8} (\bibinfo{year}{2023}).
\newblock
\urldef\tempurl%
\url{https://edisciplinas.usp.br/pluginfile.php/7614933/mod_resource/content/1/Opinion%20_%20Noam%20Chomsky_%20The%20False%20Promise%20of%20ChatGPT%20-%20The%20New%20York%20Times.pdf}
\showURL{%
\tempurl}


\bibitem[Clifton(1998)]%
        {clifton_comprehension_1998}
\bibfield{author}{\bibinfo{person}{Lyn Frazier~Charles Clifton}.} \bibinfo{year}{1998}\natexlab{}.
\newblock \showarticletitle{Comprehension of {Sluiced} {Sentences}}.
\newblock \bibinfo{journal}{\emph{Language and Cognitive Processes}} \bibinfo{volume}{13}, \bibinfo{number}{4} (\bibinfo{date}{Aug.} \bibinfo{year}{1998}), \bibinfo{pages}{499--520}.
\newblock
\showISSN{0169-0965, 1464-0732}
\urldef\tempurl%
\url{https://doi.org/10.1080/016909698386474}
\showDOI{\tempurl}


\bibitem[Cruttenden(2014)]%
        {cruttenden_gimsons_2014}
\bibfield{author}{\bibinfo{person}{Alan Cruttenden}.} \bibinfo{year}{2014}\natexlab{}.
\newblock \bibinfo{booktitle}{\emph{Gimson's pronunciation of {English}}}.
\newblock \bibinfo{publisher}{Routledge}.
\newblock
\urldef\tempurl%
\url{https://www.taylorfrancis.com/books/mono/10.4324/9780203784969/gimson-pronunciation-english-alan-cruttenden}
\showURL{%
\tempurl}


\bibitem[Hjalmarsson(2010)]%
        {hjalmarsson_human_2010}
\bibfield{author}{\bibinfo{person}{Anna Hjalmarsson}.} \bibinfo{year}{2010}\natexlab{}.
\newblock \showarticletitle{Human interaction as a model for spoken dialogue system behaviour}.
\newblock  (\bibinfo{year}{2010}).
\newblock
\urldef\tempurl%
\url{https://urn.kb.se/resolve?urn=urn:nbn:se:kth:diva-24258}
\showURL{%
\tempurl}
\newblock
\shownote{Publisher: KTH}.


\bibitem[Johnstone et~al\mbox{.}(1995)]%
        {johnstone_there_1995}
\bibfield{author}{\bibinfo{person}{Anne Johnstone}, \bibinfo{person}{Umesh Berry}, \bibinfo{person}{Tina Nguyen}, {and} \bibinfo{person}{Alan Asper}.} \bibinfo{year}{1995}\natexlab{}.
\newblock \showarticletitle{There was a long pause: influencing turn-taking behaviour in human-human and human-computer spoken dialogues}.
\newblock \bibinfo{journal}{\emph{International Journal of Human-Computer Studies}} \bibinfo{volume}{42}, \bibinfo{number}{4} (\bibinfo{date}{April} \bibinfo{year}{1995}), \bibinfo{pages}{383--411}.
\newblock
\showISSN{1071-5819}
\urldef\tempurl%
\url{https://doi.org/10.1006/ijhc.1995.1018}
\showDOI{\tempurl}


\bibitem[Jonsdottir et~al\mbox{.}(2008)]%
        {jonsdottir_learning_2008}
\bibfield{author}{\bibinfo{person}{Gudny~Ragna Jonsdottir}, \bibinfo{person}{Kristinn~R. Thorisson}, {and} \bibinfo{person}{Eric Nivel}.} \bibinfo{year}{2008}\natexlab{}.
\newblock \showarticletitle{Learning {Smooth}, {Human}-{Like} {Turntaking} in {Realtime} {Dialogue}}. In \bibinfo{booktitle}{\emph{Intelligent {Virtual} {Agents}}}, \bibfield{editor}{\bibinfo{person}{Helmut Prendinger}, \bibinfo{person}{James Lester}, {and} \bibinfo{person}{Mitsuru Ishizuka}} (Eds.). \bibinfo{publisher}{Springer}, \bibinfo{address}{Berlin, Heidelberg}, \bibinfo{pages}{162--175}.
\newblock
\showISBNx{978-3-540-85483-8}
\urldef\tempurl%
\url{https://doi.org/10.1007/978-3-540-85483-8_17}
\showDOI{\tempurl}


\bibitem[Kwiatkowski et~al\mbox{.}(2019)]%
        {kwiatkowski_natural_2019}
\bibfield{author}{\bibinfo{person}{Tom Kwiatkowski}, \bibinfo{person}{Jennimaria Palomaki}, \bibinfo{person}{Olivia Redfield}, \bibinfo{person}{Michael Collins}, \bibinfo{person}{Ankur Parikh}, \bibinfo{person}{Chris Alberti}, \bibinfo{person}{Danielle Epstein}, \bibinfo{person}{Illia Polosukhin}, \bibinfo{person}{Jacob Devlin}, \bibinfo{person}{Kenton Lee}, \bibinfo{person}{Kristina Toutanova}, \bibinfo{person}{Llion Jones}, \bibinfo{person}{Matthew Kelcey}, \bibinfo{person}{Ming-Wei Chang}, \bibinfo{person}{Andrew~M. Dai}, \bibinfo{person}{Jakob Uszkoreit}, \bibinfo{person}{Quoc Le}, {and} \bibinfo{person}{Slav Petrov}.} \bibinfo{year}{2019}\natexlab{}.
\newblock \showarticletitle{Natural {Questions}: {A} {Benchmark} for {Question} {Answering} {Research}}.
\newblock \bibinfo{journal}{\emph{Transactions of the Association for Computational Linguistics}}  \bibinfo{volume}{7} (\bibinfo{date}{Nov.} \bibinfo{year}{2019}), \bibinfo{pages}{453--466}.
\newblock
\showISSN{2307-387X}
\urldef\tempurl%
\url{https://doi.org/10.1162/tacl_a_00276}
\showDOI{\tempurl}


\bibitem[Lin and Chen(2023)]%
        {lin_llm-eval_2023}
\bibfield{author}{\bibinfo{person}{Yen-Ting Lin} {and} \bibinfo{person}{Yun-Nung Chen}.} \bibinfo{year}{2023}\natexlab{}.
\newblock \showarticletitle{{LLM}-{Eval}: {Unified} {Multi}-{Dimensional} {Automatic} {Evaluation} for {Open}-{Domain} {Conversations} with {Large} {Language} {Models}}. In \bibinfo{booktitle}{\emph{Proceedings of the 5th {Workshop} on {NLP} for {Conversational} {AI} ({NLP4ConvAI} 2023)}}, \bibfield{editor}{\bibinfo{person}{Yun-Nung Chen} {and} \bibinfo{person}{Abhinav Rastogi}} (Eds.). \bibinfo{publisher}{Association for Computational Linguistics}, \bibinfo{address}{Toronto, Canada}, \bibinfo{pages}{47--58}.
\newblock
\urldef\tempurl%
\url{https://doi.org/10.18653/v1/2023.nlp4convai-1.5}
\showDOI{\tempurl}


\bibitem[Lunsford et~al\mbox{.}(2016)]%
        {lunsford_measuring_2016}
\bibfield{author}{\bibinfo{person}{Rebecca Lunsford}, \bibinfo{person}{Peter~A. Heeman}, {and} \bibinfo{person}{Emma Rennie}.} \bibinfo{year}{2016}\natexlab{}.
\newblock \showarticletitle{Measuring {Turn}-{Taking} {Offsets} in {Human}-{Human} {Dialogues}}. In \bibinfo{booktitle}{\emph{Interspeech 2016}}. \bibinfo{publisher}{ISCA}, \bibinfo{pages}{2895--2899}.
\newblock
\urldef\tempurl%
\url{https://doi.org/10.21437/Interspeech.2016-1350}
\showDOI{\tempurl}


\bibitem[Lyu et~al\mbox{.}(2024)]%
        {lyu_real-time_2024}
\bibfield{author}{\bibinfo{person}{Ke-Ming Lyu}, \bibinfo{person}{Ren-yuan Lyu}, {and} \bibinfo{person}{Hsien-Tsung Chang}.} \bibinfo{year}{2024}\natexlab{}.
\newblock \showarticletitle{Real-time multilingual speech recognition and speaker diarization system based on {Whisper} segmentation}.
\newblock \bibinfo{journal}{\emph{PeerJ Computer Science}}  \bibinfo{volume}{10} (\bibinfo{date}{March} \bibinfo{year}{2024}), \bibinfo{pages}{e1973}.
\newblock
\showISSN{2376-5992}
\urldef\tempurl%
\url{https://doi.org/10.7717/peerj-cs.1973}
\showDOI{\tempurl}
\newblock
\shownote{Publisher: PeerJ Inc.}.


\bibitem[Magnini and Louvan(2022)]%
        {danesi_understanding_2022}
\bibfield{author}{\bibinfo{person}{Bernardo Magnini} {and} \bibinfo{person}{Samuel Louvan}.} \bibinfo{year}{2022}\natexlab{}.
\newblock \showarticletitle{Understanding {Dialogue} for {Human} {Communication}}.
\newblock In \bibinfo{booktitle}{\emph{Handbook of {Cognitive} {Mathematics}}}, \bibfield{editor}{\bibinfo{person}{Marcel Danesi}} (Ed.). \bibinfo{publisher}{Springer International Publishing}, \bibinfo{address}{Cham}, \bibinfo{pages}{1159--1201}.
\newblock
\showISBNx{978-3-031-03944-7 978-3-031-03945-4}
\urldef\tempurl%
\url{https://doi.org/10.1007/978-3-031-03945-4_20}
\showDOI{\tempurl}


\bibitem[Peng et~al\mbox{.}(2020)]%
        {peng_understanding_2020}
\bibfield{author}{\bibinfo{person}{Zhenhui Peng}, \bibinfo{person}{Kaixiang Mo}, \bibinfo{person}{Xiaogang Zhu}, \bibinfo{person}{Junlin Chen}, \bibinfo{person}{Zhijun Chen}, \bibinfo{person}{Qian Xu}, {and} \bibinfo{person}{Xiaojuan Ma}.} \bibinfo{year}{2020}\natexlab{}.
\newblock \showarticletitle{Understanding {User} {Perceptions} of {Robot}'s {Delay}, {Voice} {Quality}-{Speed} {Trade}-off and {GUI} during {Conversation}}. In \bibinfo{booktitle}{\emph{Extended {Abstracts} of the 2020 {CHI} {Conference} on {Human} {Factors} in {Computing} {Systems}}} \emph{(\bibinfo{series}{{CHI} {EA} '20})}. \bibinfo{publisher}{Association for Computing Machinery}, \bibinfo{address}{New York, NY, USA}, \bibinfo{pages}{1--8}.
\newblock
\showISBNx{978-1-4503-6819-3}
\urldef\tempurl%
\url{https://doi.org/10.1145/3334480.3382792}
\showDOI{\tempurl}


\bibitem[Piantadosi(2023)]%
        {piantadosi_modern_2023}
\bibfield{author}{\bibinfo{person}{Steven Piantadosi}.} \bibinfo{year}{2023}\natexlab{}.
\newblock \showarticletitle{Modern language models refute {Chomsky}’s approach to language}.
\newblock \bibinfo{journal}{\emph{Lingbuzz Preprint, lingbuzz}}  \bibinfo{volume}{7180} (\bibinfo{year}{2023}).
\newblock
\urldef\tempurl%
\url{https://lingbuzz.net/lingbuzz/007180/current.pdf}
\showURL{%
\tempurl}


\bibitem[Reimers and Gurevych(2019)]%
        {reimers_sentence-bert_2019}
\bibfield{author}{\bibinfo{person}{Nils Reimers} {and} \bibinfo{person}{Iryna Gurevych}.} \bibinfo{year}{2019}\natexlab{}.
\newblock \bibinfo{title}{Sentence-{BERT}: {Sentence} {Embeddings} using {Siamese} {BERT}-{Networks}}.
\newblock
\newblock
\urldef\tempurl%
\url{http://arxiv.org/abs/1908.10084}
\showURL{%
\tempurl}
\newblock
\shownote{arXiv:1908.10084 [cs]}.


\bibitem[Rosset et~al\mbox{.}(2024)]%
        {rosset_researchy_2024}
\bibfield{author}{\bibinfo{person}{Corby Rosset}, \bibinfo{person}{Ho-Lam Chung}, \bibinfo{person}{Guanghui Qin}, \bibinfo{person}{Ethan~C. Chau}, \bibinfo{person}{Zhuo Feng}, \bibinfo{person}{Ahmed Awadallah}, \bibinfo{person}{Jennifer Neville}, {and} \bibinfo{person}{Nikhil Rao}.} \bibinfo{year}{2024}\natexlab{}.
\newblock \bibinfo{title}{Researchy {Questions}: {A} {Dataset} of {Multi}-{Perspective}, {Decompositional} {Questions} for {LLM} {Web} {Agents}}.
\newblock
\newblock
\urldef\tempurl%
\url{https://doi.org/10.48550/arXiv.2402.17896}
\showDOI{\tempurl}
\newblock
\shownote{arXiv:2402.17896 [cs]}.


\bibitem[Stivers et~al\mbox{.}(2009)]%
        {stivers_universals_2009}
\bibfield{author}{\bibinfo{person}{Tanya Stivers}, \bibinfo{person}{N.~J. Enfield}, \bibinfo{person}{Penelope Brown}, \bibinfo{person}{Christina Englert}, \bibinfo{person}{Makoto Hayashi}, \bibinfo{person}{Trine Heinemann}, \bibinfo{person}{Gertie Hoymann}, \bibinfo{person}{Federico Rossano}, \bibinfo{person}{Jan~Peter de Ruiter}, \bibinfo{person}{Kyung-Eun Yoon}, {and} \bibinfo{person}{Stephen~C. Levinson}.} \bibinfo{year}{2009}\natexlab{}.
\newblock \showarticletitle{Universals and cultural variation in turn-taking in conversation}.
\newblock \bibinfo{journal}{\emph{Proceedings of the National Academy of Sciences of the United States of America}} \bibinfo{volume}{106}, \bibinfo{number}{26} (\bibinfo{date}{June} \bibinfo{year}{2009}), \bibinfo{pages}{10587--10592}.
\newblock
\showISSN{0027-8424}
\urldef\tempurl%
\url{https://doi.org/10.1073/pnas.0903616106}
\showDOI{\tempurl}


\bibitem[Zheng et~al\mbox{.}(2023)]%
        {zheng_judging_2023}
\bibfield{author}{\bibinfo{person}{Lianmin Zheng}, \bibinfo{person}{Wei-Lin Chiang}, \bibinfo{person}{Ying Sheng}, \bibinfo{person}{Siyuan Zhuang}, \bibinfo{person}{Zhanghao Wu}, \bibinfo{person}{Yonghao Zhuang}, \bibinfo{person}{Zi Lin}, \bibinfo{person}{Zhuohan Li}, \bibinfo{person}{Dacheng Li}, \bibinfo{person}{Eric~P. Xing}, \bibinfo{person}{Hao Zhang}, \bibinfo{person}{Joseph~E. Gonzalez}, {and} \bibinfo{person}{Ion Stoica}.} \bibinfo{year}{2023}\natexlab{}.
\newblock \bibinfo{title}{Judging {LLM}-as-a-{Judge} with {MT}-{Bench} and {Chatbot} {Arena}}.
\newblock
\newblock
\urldef\tempurl%
\url{https://doi.org/10.48550/arXiv.2306.05685}
\showDOI{\tempurl}
\newblock
\shownote{arXiv:2306.05685 [cs]}.


\bibitem[Zhou et~al\mbox{.}(2022)]%
        {zhou_online_2022}
\bibfield{author}{\bibinfo{person}{Jiawei Zhou}, \bibinfo{person}{Jason Eisner}, \bibinfo{person}{Michael Newman}, \bibinfo{person}{Emmanouil~Antonios Platanios}, {and} \bibinfo{person}{Sam Thomson}.} \bibinfo{year}{2022}\natexlab{}.
\newblock \showarticletitle{Online {Semantic} {Parsing} for {Latency} {Reduction} in {Task}-{Oriented} {Dialogue}}. In \bibinfo{booktitle}{\emph{Proceedings of the 60th {Annual} {Meeting} of the {Association} for {Computational} {Linguistics} ({Volume} 1: {Long} {Papers})}}, \bibfield{editor}{\bibinfo{person}{Smaranda Muresan}, \bibinfo{person}{Preslav Nakov}, {and} \bibinfo{person}{Aline Villavicencio}} (Eds.). \bibinfo{publisher}{Association for Computational Linguistics}, \bibinfo{address}{Dublin, Ireland}, \bibinfo{pages}{1554--1576}.
\newblock
\urldef\tempurl%
\url{https://doi.org/10.18653/v1/2022.acl-long.110}
\showDOI{\tempurl}


\end{thebibliography}

\end{document}